\documentclass[12pt, a4paper]{article}       
\usepackage{amssymb}
\usepackage{graphicx}
\begin{document} 
\begin{center}
{\large \bf Central and peripheral interactions of hadrons}

\vspace{0.5cm}                   

{\bf I.M. Dremin$^{1,2}$, V.A. Nechitailo$^{1}$, S.N. White$^3$}

\vspace{0.5cm}                       

        $^1$Lebedev Physics Institute, Moscow 119991, Russia

\medskip

    $^2$National Research Nuclear University "MEPhI", Moscow 115409, Russia     

\medskip 

   $^3$CERN, CH-1211 Geneva 23, Switzerland

\end{center}

\begin{abstract}
 Surprisingly enough, the ratio of elastic to inelastic cross sections
of proton interactions increases with energy in the interval 
corresponding to ISR$\rightarrow $LHC (i.e. from 10 GeV to 10$^4$ GeV).
That leads to special features of their spatial interaction 
region at these and higher energies. Within the framework of
some phenomenological models, we show how the particular ranges of the 
transferred momenta measured in elastic scattering experiments expose 
the spatial features of the inelastic interaction region according to the unitarity 
condition. The difference between their predictions at higher energies is 
discussed. The notion of central and peripheral collisions of hadrons is treated 
in terms of the impact parameters description. It is shown that the shape of the 
differential cross section in the diffraction cone is mostly determined 
by collisions with intermediate impact parameters. Elastic scattering 
at very small transferred momenta is sensitive to peripheral processes with 
large impact parameters. The role of central collisions in formation of the 
diffraction cone is less significant.
\end{abstract}

\section{Introduction}

Traditionally, hadron collisions were classified according to our
prejudices about hadron structure. From the earlier days of Yukawa's prediction
of pions, the spatial extent of hadrons was ascribed in pre-QCD times to the 
pion cloud (of size of the inverse pion mass) surrounding their centers. 
The very external shell was described as formed by single virtual pions 
representing the lightest particle constituents. The deeper shells were 
occupied by heavier objects (2$\pi , \rho $-mesons etc.). 
In the quantum field theory, these objects contribute to hadron 
scattering amplitudes by their propagators polynomially damped for
transferred momenta of the order of the corresponding masses. Therefore, according 
to the Heisenberg principle, the largest spatial extent is typical for the 
single pion (smallest masses!) exchange. That is why the one pion exchange 
model was first proposed \cite{drch} for the description of inelastic peripheral 
interactions of hadrons. Later on, more central collisions with the exchange of 
$\rho $-mesons and all other heavier Regge
particles were considered and included in the multiperipheral models.

Some knowledge about the spatial extent of inelastic hadron interactions 
can also be gained from properties of their elastic scattering connected with 
inelastic processes by the unitarity condition. The spatial view of the 
collision process of two hadrons is not directly observable because of their 
extremely small sizes and its short time-duration. However it is very important 
for our heuristic view. The interaction region is characterized by the impact 
parameter $b$ which denotes the shortest transverse distance between the 
trajectories of the centers of the colliding hadrons. Different spatial regions
are responsible for the relation with different ranges of the transferred 
momenta in the experimentally measured differential cross secton.

To analyze this relation we choose the particular QCD-motivated model 
\cite{kfk1, kfk2} which we call by the first letters of the names of its 
coauthors as the kfk-model. This model has described quite 
precisely the present experimental measurements of the elastic scattering of 
protons from ISR to LHC energies. The energy dependence imposed by the model is 
used for predictions at higher energies. We compare its conclusions with other 
approaches to the problem.

The kfk-model provides analytically the shapes of the elastic scattering 
amplitude in terms of both the transferred momenta $t$ measured 
experimentally and the impact parameters $b$ relevant for the spatial view
of the process. That allows one to study separately different regions of them 
and their mutual influence, i.e. to reveal "the anatomy" of the model. 

We consider both approaches and show: 
\begin{enumerate}
 \item How different $t$-regions represented by the measured differential cross
   section contribute to the shape of the spatial interaction region and to 
   the unitarity condition (sections 5 and 6); 
\item How different spatial $b$-regions contribute to the measurable $t$-structure 
   of the elastic scattering amplitude (section 7).         
\end{enumerate}

\section {The kfk-model}

The kfk-model \cite{kfk1, kfk2} originates from the so-called  Stochastic Vacuum 
Model \cite{dos, dfk} which was initially formulated by deriving the impact
parameter shape of the elastic scattering amplitude using some QCD-motivated
arguments. Applying the Fourier - Bessel transformation one gets the elastic
amplitude in terms of the transferred momenta. It has been shown \cite{fp, kfk1}
that the resulting shape describes well the data on 
$d\sigma /dt, \sigma _{el}, \sigma _{tot}$ at energies from ISR (11 - 60 GeV
in the center of mass system) \cite{amal} through LHC (2.76  - 13 TeV)
\cite{tot1, tot2, tot3, tot4, cs, atl1, atl2} with the help of the definite set
of the energy dependent parameters.

The differential cross section is defined as 
\begin{equation}
d\sigma /dt =\vert f(s,t)\vert ^2=f^2_I+f^2_R,
\label{dsdt}
\end{equation}
where the labels $K=I,R$ denote correspodingly the imaginary and real parts
of the elastic amplitude $f(s,t)$ (of the dimension GeV$^{-2}$). The variables
$s$ and $t$ are the squared energy and transferred momentum of 
colliding protons in the center of mass system $s=4E^2=4(p^2+m^2)$, 
$-t=2p^2(1-\cos \theta)$ at the scattering angle $\theta $.  

The nuclear part of the amplitude $f$ in the kfk-model is
\begin{equation}
f_K(s,t)= \alpha _K(s)e^{-\beta _K\vert t\vert}+
\lambda _K(s)\Psi _K(\gamma _K(s),t),
\label{fk}
\end{equation}
with the characteristic shape function
\begin{equation}
\Psi _K(\gamma _K(s),t)=2e^{\gamma _K}\left[
\frac {e^{-\gamma _K\sqrt {1+a_0\vert t\vert}}}{\sqrt {1+a_0\vert t\vert}} -
e^{\gamma _K}\frac {e^{-\gamma _K\sqrt {4+a_0\vert t\vert}}}
{\sqrt {4+a_0\vert t\vert}}\right]. 
\label{psik}
\end{equation}
In what follows, we use the explicit expressions for the energy dependent
parameters $\alpha, \beta, \gamma, \lambda $ shown in \cite{kfk1} which fitted
the data. In total, there are 8 such parameters each of which contains the 
energy independent terms and those increasing with energy $s$ as $\log \sqrt s$ 
and $\log ^2\sqrt s$ (see Eqs (29)-(36) in \cite{kfk2}). Thus 8
coefficients have been determined from comparison with experiment at a given
energy and 24 for the description of the energy dependence in a chosen interval. 
The parameter $a_0=1.39$ GeV$^{-2}$ is proclaimed to be fixed.

Let us note that we have omitted the nuclear-Coulomb interference term because
of the extremely tiny region of small transferred momenta where it becomes 
noticeable. 

The corresponding dimensionless nuclear amplitude in the $b$-representation 
is written as
\begin{equation}
\tilde{f}_K(s,b)= \frac {\alpha _K}{2\beta _K}e^{-b^2/4\alpha _K}+
\lambda _K \tilde{\Psi }_K(s,b)
\label{fkb}
\end{equation}
with
\begin{equation}
\tilde{\Psi }_K(s,b)=\frac {2e^{\gamma _K-\sqrt {\gamma ^2_K +b^2/a_0}}}
{a_0\sqrt {\gamma ^2_K +b^2/a_0}}(1-e^{\gamma _K-\sqrt {\gamma ^2_K +b^2/a_0}}).
\label{psib}
\end{equation}
The two-dimensional Fourier transformation used is
\begin{equation}
\tilde{f}(s,b)=\frac {1}{2\pi }\int d^2{\bf q}e^{-i{\bf q}{\bf b}}f(s,t=-q^2).
\label{fsbt}
\end{equation}

\section{The unitarity condition}

From the theoretical side, the most reliable (albeit rather limited) information 
about the relation between elastic and inelastic processes comes from the 
unitarity condition. The unitarity of the 
$S$-matrix $SS^+$=1 relates the amplitude of elastic
scattering $f(s,t)$ to the amplitudes of inelastic processes $M_n$. 
In the $s$-channel they are subject to the integral relation (for more
details see, e.g., \cite{PDG, ufnel1, ufnel2, ufnel3}) which can be written 
symbolically as
\begin{equation}
f_I(s,t)= I_2(s,t)+g(s,t)=
\int d\Phi _2 f(s,t_1)f^*(s,t_2)+\sum _n\int d\Phi _n M_nM_n^*.
\label{unit}
\end{equation}
The non-linear integral term represents the two-particle intermediate states of 
the incoming particles integrated over transferred momenta $t_1$ and $t_2$ 
combining to final $t$. The second term represents the shadowing contribution of 
inelastic processes to the imaginary part of the elastic scattering 
amplitude. Following \cite{hove} it is called the overlap function. This 
terminology is ascribed to it because the integral there defines the overlap 
within the corresponding phase space $d\Phi _n$ between the matrix element 
$M_n$ of the $n$-th inelastic channel and its conjugated counterpart with the 
collision axis of initial particles deflected by an angle $\theta $ in proton 
elastic scattering. It is positive at $\theta =0$ but can 
change sign at $\theta \neq 0$ due to the relative phases of inelastic matrix 
elements $M_n$'s.

At $t=0$ it leads to the optical theorem 
\begin{equation}
f_I(s,0)=\sigma _{tot}/4\sqrt {\pi}
\label{opt}
\end{equation}
and to the general statement that the total cross section is the sum of 
cross sections of elastic and inelastic processes
\begin{equation}
\sigma _{tot}=\sigma _{el}+\sigma _{inel}.
\label{telin}
\end{equation}
If divided by $\sigma _{tot}$ this relation tells
that the total probability of all processes is equal to one.

It is possible to study the space structure of the 
interaction region of colliding protons using information about their 
elastic scattering within the framework of the unitarity condition. The whole procedure is
simplified because in the space representation one gets an algebraic
relation between the elastic and inelastic contributions to the unitarity
condition in place of the more complicated non-linear integral term 
$I_2$ in Eq. (\ref{unit}).

In what follows we prefer to use the different notation of $\tilde{f}$ 
in (\ref{fsbt}) for clearer distinction from $f$:
\begin{equation}
i\Gamma (s,b)=\frac {1}{2\sqrt {\pi }}\int _0^{\infty}d\vert t\vert f(s,t)
J_0(b\sqrt {\vert t\vert }).
\label{gamm}
\end{equation}
Applying directly the transformation (\ref{gamm}) to the relation (\ref{unit})
one gets the unitarity condition in the $b$-representation
\begin{equation}
G(s,b)=2\Gamma _R(s,b)-\vert \Gamma (s,b)\vert ^2.
\label{unit1}
\end{equation}
The left-hand side (the overlap function in the $b$-representation) describes 
the transverse impact-parameter profile of inelastic collisions of protons. It 
is just the Fourier -- Bessel transform of the overlap function $g$. It 
satisfies the inequalities $0\leq G(s,b)\leq 1$ and determines how absorptive 
the interaction region is, depending on the impact parameter (with larger $G$ 
for larger absorption). The profile of elastic processes is determined by the 
subtrahend in Eq. (\ref{unit1}). Thus Eq. (\ref{unit1}) establishes the 
relation between the elastic and inelastic impact-parameter distributions
$G(s,b=d^2\sigma _{inel}/db^2$ and $\vert \Gamma (s,b)\vert ^2=
d^2\sigma _{el}/db^2$ with $2\Gamma _R(s,b)=d^2\sigma _{tot}/db^2$. 

If $G(s,b)$ is
integrated over all impact parameters, it leads to the cross section for 
inelastic processes. The terms on the right-hand side would produce the total
cross section and the elastic cross section, correspondingly, as should be the 
case according to Eq. (\ref{telin}). The overlap 
function is often discussed in relation with the opacity (or the eikonal phase) 
$\Omega (s,b)$ such that $G(s,b)=1-\exp (-\Omega (s,b))$. Thus, larger
absorption corresponds to larger $\Omega $. 

\section{Brief review of the elastic scattering data}

The Eq. (\ref{unit1}) shows that the inelastic profile $G$ is directly
expressed in terms of the elastic amplitude. Therefore let us describe
experimentally measured properties of elastic scattering and discuss how they 
are reproduced by the kfk-model. 

The bulk features of the differential cross section at high energies with
increase of the transferred momentum $\vert t\vert$ can be briefly stated 
as its fast decrease at low transferred momenta within the 
diffraction cone and somewhat slower decrease at larger momenta. The 
diffraction cone is usually approximated by the exponent $\exp (B(s)t)$ 
while further decrease in the so-called Orear region is roughly fitted by 
a smaller than $B$ slope or by the dependence of the type 
$\exp (-c\sqrt {\vert t\vert})$. For the recent data at 13 TeV see 
\cite{csorg}.

There are some 
special features noted. The increase towards $t=0$ in the very tiny region 
of extremely low momenta becomes steeper. That is ascribed to the interference 
of nuclear and Coulomb amplitudes. It serves to determine the ratio of the 
real and imaginary parts of the amplitude $\rho =f_R/f_I$ at $t=0$. In the 
transition region between the two main regimes the differential
cross section flattens somewhat and/or shows some dip. It will be specially
discussed below. At transferred momenta larger than those of the Orear region
further flattening is observed. This tail is usually described perturbatively
by the three-gluon exchange with real amplitude. However the cross section is 
so small there that this region is unimportant for our conclusions.

The optical theorem (\ref{opt}) assures us that the imaginary part of the
amplitude in forward direction must be positive at all energies. The real part 
at $t=0$ has been measured also to be positive at high energies and comparatively
small ($\rho \approx 0.1 - 0.14$). For recent data at 13 TeV see \cite{gia}.
This result agrees with predictions of the 
dispersion relations. Thus it only contributes about 1$\%$ or 2$\%$ to the 
forward differential cross section (\ref{dsdt}). 

In principle, both real and imaginary parts can be as positive as negative
at other transferred momenta. Anyway, they are bounded by 
the values $\pm \sqrt {d\sigma /dt}$ and must be small in those $t$-regions 
where the differential cross section is small. Actually, these two bounds
are used for two different approaches considered below. They determine the 
difference between their predictions about the shape of the interaction region.

The further knowledge about the elastic amplitudes comes either from some 
theoretical considerations or from general guesses and the model building. 
It was shown in papers \cite{mar1, mar2} that the real part of the amplitude 
must change its sign somewhere in the diffraction cone. Therefore, its decrease
with increasing $\vert t\vert $ inside the diffraction cone
must be faster than that of the imaginary part which then should mainly 
determine the value of the slope $B(s)$. The dip between the two main
typical regimes of the diffraction cone and Orear behavior
 inspires the speculation that the imaginary part also passes
through zero near the dip. Then the dip of the differential cross section 
is filled in by the small real part of the amplitude. These guesses are well 
supported by the results of the kfk-model used by us.

\section{How different $t$-regions contribute to the $b$-shape of the 
interaction region}

At the outset we will not discuss the spatial extension of the interaction 
region as a function of the impact parameter $b$. It was carefully studied in 
several publications \cite{amal, dnec, ijmp, igse, white, mart}. Instead, we 
limit ourselves by the simpler case of the energy dependence of the intensity of 
interaction for central (head-on) collisions of impinging protons at $b=0$. 
That is the most sensitive point of the whole picture demonstrating its
crucial features.

Let us introduce the variable $\zeta $:
\begin{equation}
\zeta (s)=\Gamma _R(s,0)=\frac {1}{2\sqrt {\pi }}\int _0^{\infty}
d\vert t\vert f_I(s,t).
\label{zeta1}
\end{equation}         
To compute the integral, one must know the behavior of the imaginary part of
the amplitude for all transferred momenta $t$ at a given energy $s$. The choice
of the sign of $f_I$ at large $\vert t\vert $ is important for model conclusions.

Now, have a look at the second term in the unitarity condition (\ref{unit1}):
\begin{equation}
\vert \Gamma (s,0)\vert ^2=\zeta ^2+\frac {1}{4\pi}\left(\int _0^{\infty}
d\vert t\vert f_R(s,t)\right) ^2.
\label{g2}       
\end{equation}
The last term here can be neglected compared to the first one. That is easily
seen from the inequalities 
\begin{equation}
2\sqrt {\pi}\Gamma _I(s,0)=\int _0^{\infty}d\vert t\vert f_R \leq
\int _0^{\infty}d\vert t\vert \vert f_R\vert =
\int _0^{\infty}d\vert t\vert \sqrt 
{\frac {\rho ^2(s,t)d\sigma /dt}{1+\rho ^2(s,t)}}\ll 2\sqrt {\pi}\zeta
\label{g3}       
\end{equation}
Here $\rho (s,t)=f_R(s,t)/f_I(s,t).$ The factor $\rho ^2(s,t)/(1+\rho ^2(s,t))$ 
is very small in the diffraction cone because $\rho ^2(s,0)\leq 0.02$ 
according to experimental results and $\rho (s,t)$ should possess zero 
inside the diffraction cone \cite{mar1, mar2}. It can become of the order 1 at 
large values of $\rho ^2(s,t)$ (say, at the dip) but the cross section 
$d\sigma /dt$ is small there already \cite{tot1, anddre1}. The smallness of the
contribution from the real part to $\vert \Gamma \vert ^2$ is strongly supported
by the kfk-model as shown in the Table below.

Then the unitarity condition (\ref{unit1}) for central collisions can be 
written as 
\begin{equation}
G(s,b=0)= \zeta (s) (2-\zeta (s)).
\label{gZ}       
\end{equation}
Thus, according to the unitarity condition (\ref{gZ}) the darkness $G(s,0)$
of the inelastic interaction region for central collisions (absorption) is 
defined by the single energy dependent parameter $\zeta (s)$. 
It has the maximum $G(s,0)=1$ for $\zeta =1$. Any decline of $\zeta $ from 1
($\zeta =1\pm \epsilon $) results in the parabolic decrease of the absorption
($G(s,0)=1-\epsilon ^2$), i.e. in an even much smaller decline from 1 for
small $\epsilon $. The elastic profile, equal to $\zeta ^2$ in central 
collisions, also reaches the value 1 for $\zeta =1$. Namely the point $b=0$
happens to be most sensitive to the variations of $\zeta $ in different models.

Formally, the unitarity condition (\ref{gZ}) imposes the limit $\zeta \leq 2$. 
It is required by the positivity of the inelastic profile. This limit 
corresponds to the widely discussed "black disk" picture which asks for the 
relation
\begin{equation}
 \sigma _{el}=\sigma _{inel}=\sigma _{tot}/2.
\end{equation}

\begin{figure}
\caption{The experimental data about the 
diffraction cone slope $B(s)$ \cite{tot5}.}
\includegraphics[width=\textwidth]{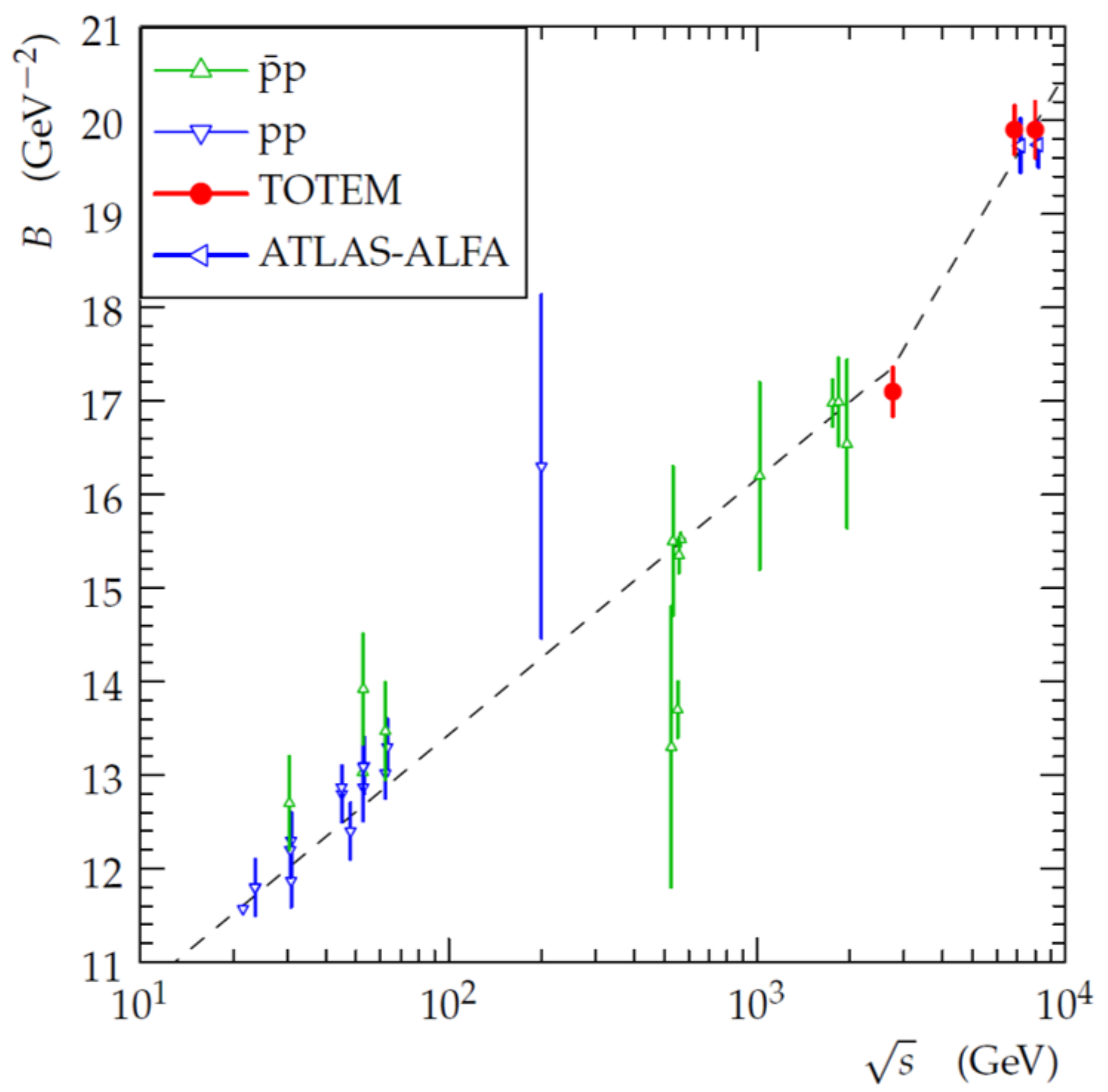}
\label{fig:TOTEM_B(s)}
\end{figure}

\begin{figure}
\caption{The fit of the $B$-data (Fig.1) by the kfk-model (the dotted line). 
The dashed line $B_I$ indicates that the slope is well accounted by the 
imaginary part of the amplitude alone.}
\includegraphics[width=\textwidth]{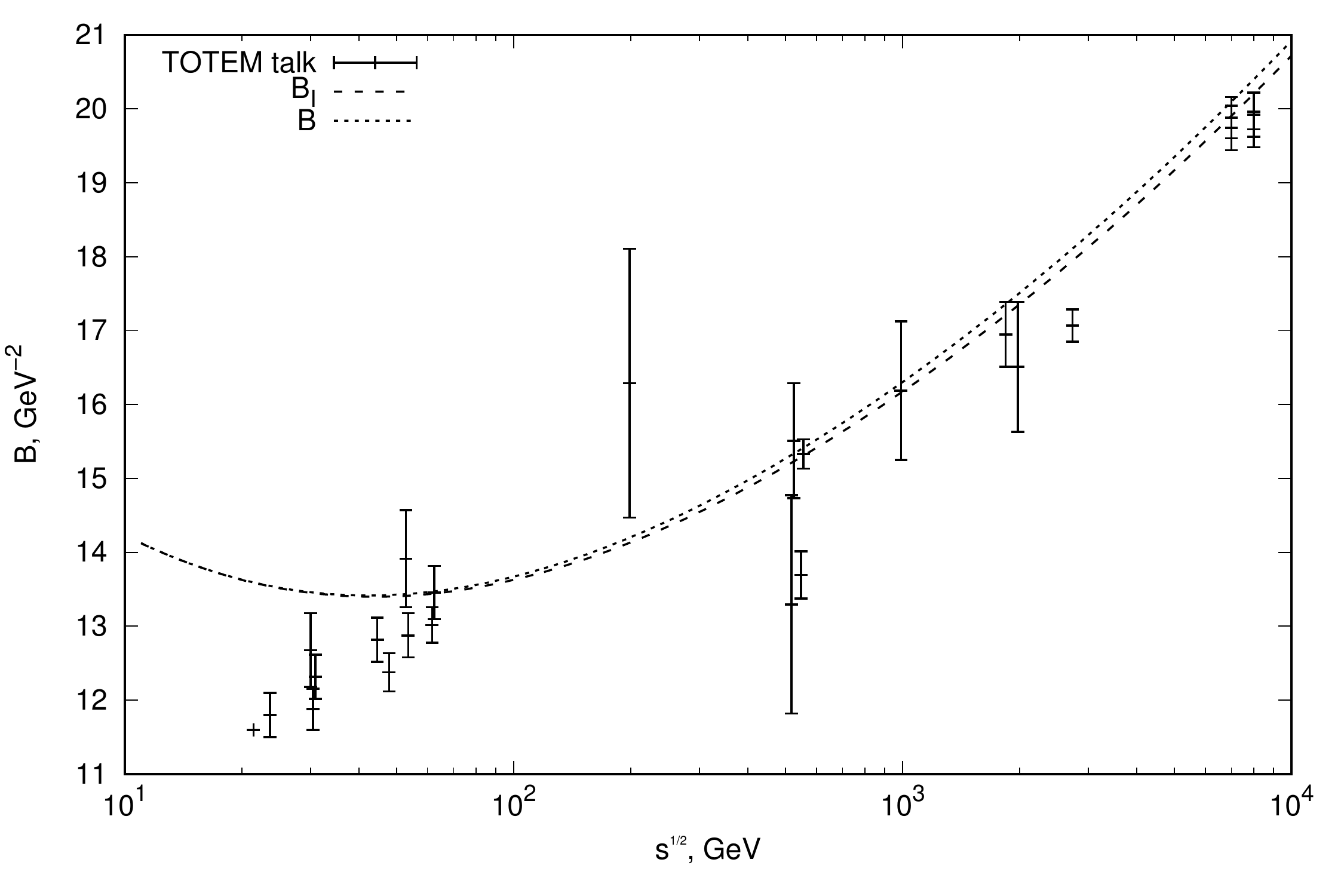}
\label{fig:B(s)}
\end{figure}

Both real and imaginary parts of the amplitude are analytically prescribed by 
the kfk-model. That allows one to calculate the experimental characteristics and 
get direct insight into the validity of some approximations. 
Figure~\ref{fig:TOTEM_B(s)} reproduces the experimental data about the 
diffraction cone slope $B(s)$ \cite{tot5} and Figure~\ref{fig:B(s)} demonstrates
how they are fitted by the kfk-model from ISR to LHC energies. The fit is good
in general but one sees some discrepancy at low ISR energies and at TOTEM 
2.76 TeV data. 

In the Table we show the energy 
dependence of several characteristics of proton collisions with estimates of 
the role of different regions of integration in (\ref{zeta1}) when computed 
according to the prescriptions for $f_I$ and $f_R$ of the kfk-model.
This "anatomy" answers the question raised at the title of the section.

\medskip
\begin{table}
\caption{The energy behavior of main $pp$-characteristics of the kfk-model.
(Their detailed explanation is given in the text).}
\begin{tabular}{|l|l|l|l|l}
        \hline
$\sqrt s$, TeV                     & 7         &  100      & 10$^4$  \\ \hline
$\zeta $                           &  0.95058  &  0.99531  &  0.99793  \\
$\vert t_0\vert ,$GeV$^2$          &  0.47566  &  0.28458  &  0.13978 \\
$\zeta [0,\vert t_0\vert ]$        &  0.99193  &  1.0769   &  1.1505 \\ 
$\zeta [\vert t_0\vert,\infty ]$   & -0.041349 & -0.081548 & -0.15260\\ 
$G(s,0)$                           &  0.99756  &  0.99998  &  1.00000 \\
$\Gamma _I[0,\vert t_0\vert]$      &  0.053928 &  0.054432 &  0.047289  \\ 
$\Gamma _I[\vert t_0\vert,\infty ]$&  0.003405 &  0.001024 & -0.001311  \\ 
$\Gamma ^2_I$                      &  0.003287 &  0.003075 &  0.002114 \\                     
$4\sigma _{el}/\sigma _{tot}$      &  1.03     &  1.148    &  1.26 \\
$\sigma _{inel}/\sigma _{el}$      &  2.89     &  2.48     &  2.17 \\
$B\vert t_0\vert$                  &  9.4676   &  7.6956   &  6.2829\\ \hline
$\zeta (\sigma)$                   & 1.0333    &  1.1584   &  1.3031 \\ \hline
\end{tabular}
\end{table}

\begin{figure}
\caption{The $t$-dependence of $f_I$ at 7 TeV and at 10$^4$ TeV (the kfk-model). 
 The positions of zeros are shown.}
\includegraphics[width=\textwidth]{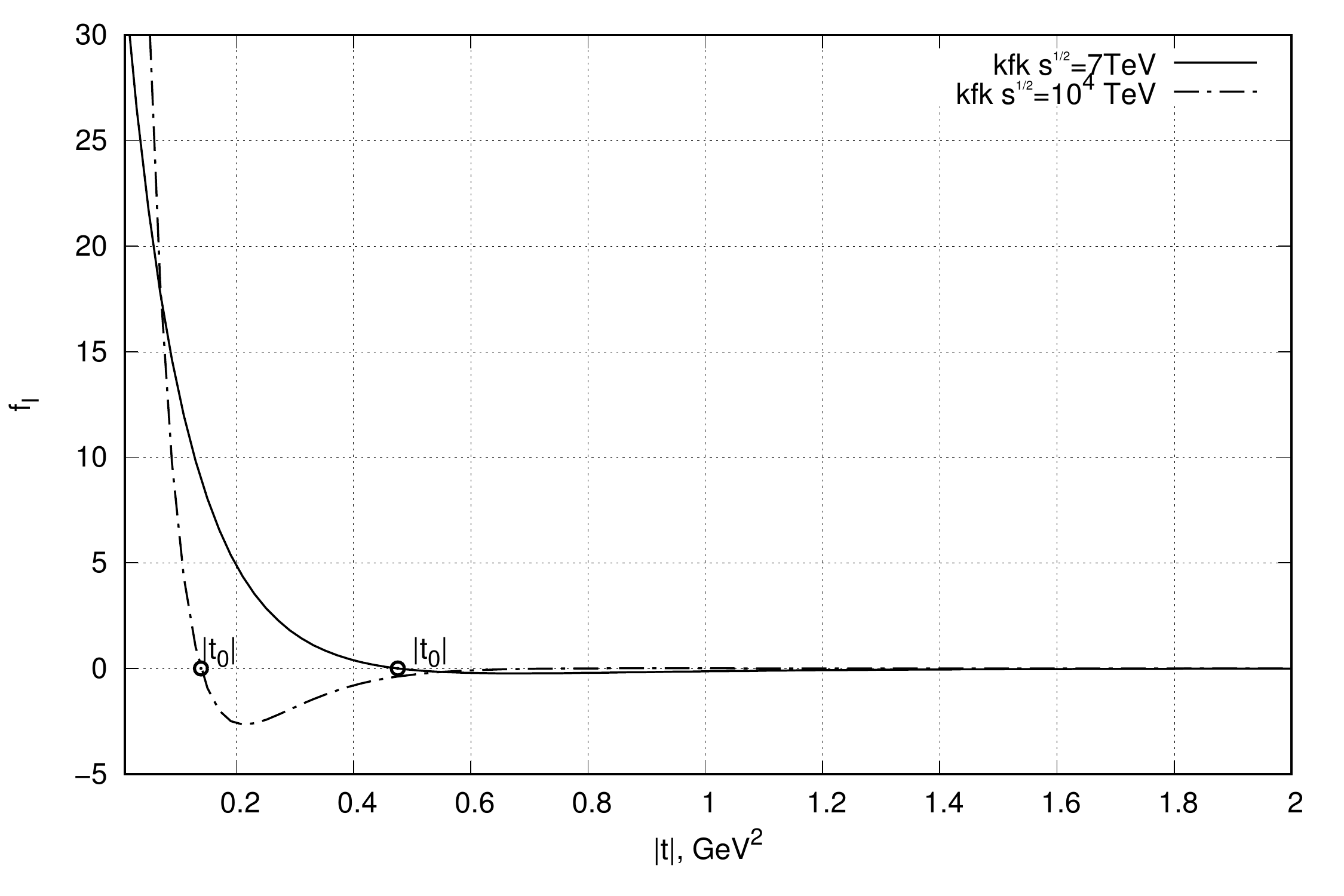}
\label{fig:TI}
\end{figure}

The exact value of $\zeta (s)$ is crucial for our conclusions, especially 
if $\zeta $ closely approaches 1 as occurs at LHC energies. Its values
in the kfk-model at energies from 7 TeV to 10$^4$ TeV are shown in first line
of the Table. They approach 1 asymptotically from below. The integrand 
$f_I$ in Eq. (\ref{zeta1}) changes sign for the kfk-model as demonstrated at 7 
TeV and 10$^4$ TeV in Figure~\ref{fig:TI}. The energy dependence of its zeros 
$t=t_0(s)$ has been computed and shown in the Table. The contributions to 
$\zeta $ from the positive and negative branches of the integrand are also shown 
there as $\zeta [0,\vert t_0\vert]$ and $\zeta [\vert t_0\vert,\infty]$.
The first of them exceeds 1 at high energies and would deplete the inelastic
profile at the center if not compensated by the second one.
The numerical contribution of the negative tail is very small but it is
decisive for the asymptotic behavior of $\zeta $ which approaches 1 and does
not exceed it. Thus the central profile saturates with $G(s,0)$ tending to 1
(see line 5 of the Table). The whole profile becomes more Black, Edgier and 
Larger (the BEL-regime \cite{rhpv}) similar to the tendency observed from ISR 
to LHC energies (see Fig.~7 in \cite{kfk2}). Nothing drastic happens in 
asymptopia! 

The share of elastic processes and their ratio to inelastic collisions computed 
according to the kfk-model are also shown in the Table (lines 9 and 10). Let us 
stress again that in the kfk-model the ratio $r=4\sigma _{el}/\sigma _{tot}$ 
becomes larger than 1 with increasing energy (line 9) while $\zeta $ saturates 
at 1 as shown in the Table (line 1). This difference is crucial for asymptotic 
predictions about the shape of the interaction region.

To support our assumption that the real part can be neglected according to
estimates of Eq. (\ref{g3}), we have directly computed 
$\Gamma _I[0,\vert t_0\vert ]=
\int _0^{\vert t_0\vert }d\vert t\vert f_R(s,t)/2\sqrt {\pi}$ and 
$\Gamma _I[\vert t_0\vert, \infty]=
\int _{\vert t_0\vert }^{\infty}d\vert t\vert f_R(s,t)/2\sqrt {\pi }$ for the 
kfk-model. In accordance with Eq. (\ref{g3}) they happen to be negligibly small 
as seen from the Table at $\Gamma _I[...]$- and $\Gamma _I^2$-lines.

\begin{figure}
\caption{The value of $B\vert t_0\vert $ decreases with energy (the kfk-model).}
\includegraphics[width=\textwidth]{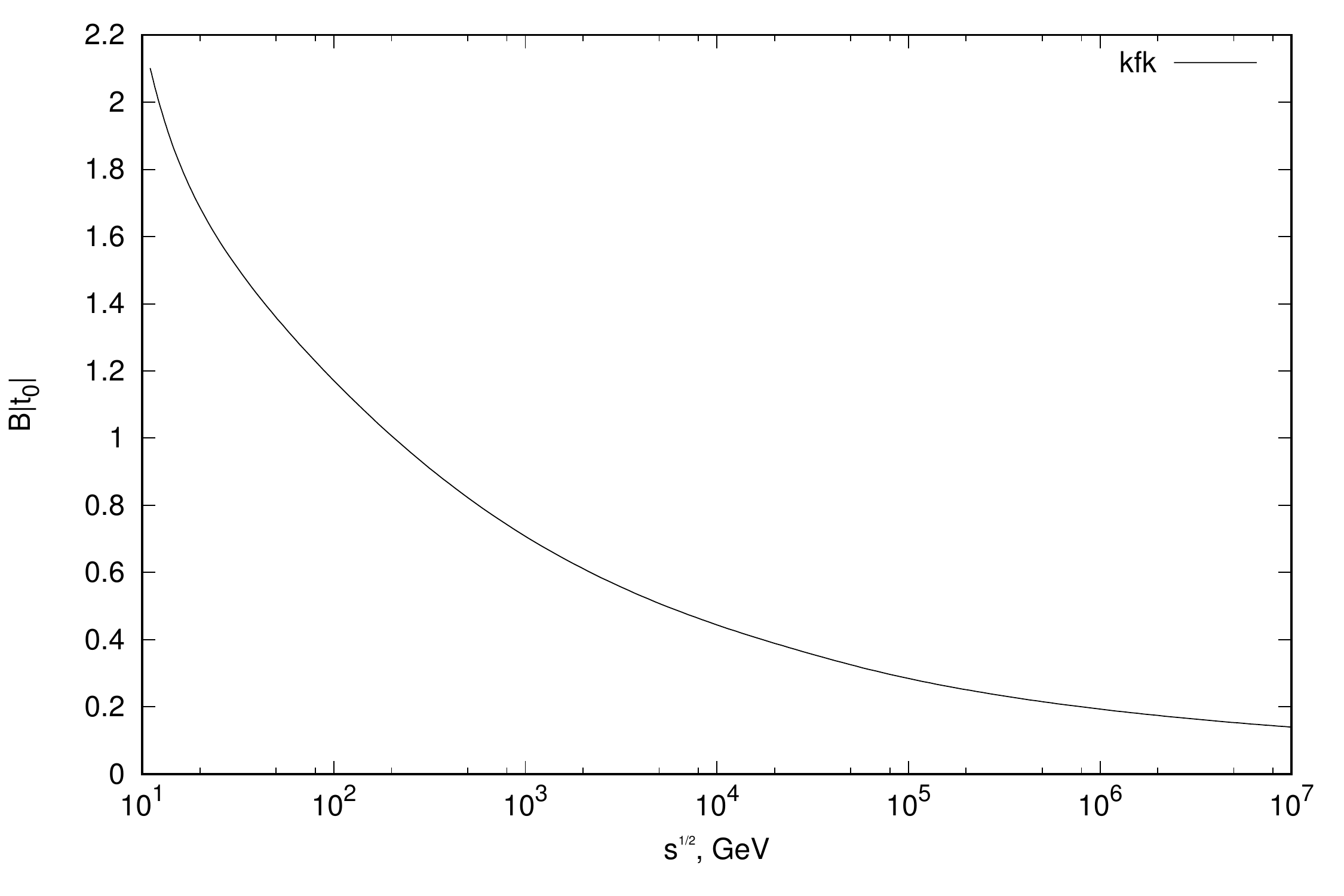}
\label{fig:Bt0}
\end{figure}

It is interesting to note, that, according to the kfk-model, the contribution 
of the imaginary part to the differential cross section dominates almost 
everywhere except the very narrow region near the dip (see Fig.~3 
in \cite{kfk2}). This is also true at those transferred momenta where 
$f_I$ becomes negative. Surely, the real part dominates in the narrow dip
region of the differential cross section near $t_0$ but its integral 
contribution to $\vert \Gamma \vert ^2$ can be neglected (compare line 1
and line 8 of the Table). 

We do not consider in detail the whole impact-parameter shape of the 
interaction region here because, for our purposes, it was enough to consider
it at the most sensitive point of central collisions at $b=0$. Moreover, 
it has been done in the references \cite{amal, dnec, ijmp, igse, white, mart}. 

\section{Comparison to some other models}

There is no shortage of models of elastic scattering on the market
nowadays. We want just to stress that the kfk-model provides explicit 
analytical expressions for the amplitude both in transferred momenta and
in impact parameters.

The $t$-behavior of their amplitude was compared by the authors with the 
predictions of the BSW-model \cite{bsw} in Refs \cite{kfk1, kfk2} and 
with Selyugin-model \cite{sel} in Ref. \cite{kfk2} at energies 7 and 14 TeV.
In the measured interval of the transferred momenta they almost coincide
within the experimental uncertainties. The results of the models are slightly 
different in the shapes at larger transferred momenta due to the different 
number of predicted zeros of the amplitude and can be compared with newly
measured data \cite{csorg, gia}.

The two most important features of the experimental data are well reproduced 
by the kfk-model. Those are the slope of the diffraction cone and the dip 
position which coincides practically with the zero of the imaginary part of the 
amplitude. Therefore, to get the easier insight, one can oversimplify the model 
leaving only these two parameters \cite{sur} and writing 
$f_I\propto (1-(t/t_0)^2)\exp (Bt/2)$ with the slope 
$B=(B_I+\rho ^2B_R)/(1+\rho ^2)\approx B_I$. This toy-model admits to
calculate analytically the relation between $r$ and $\zeta $:
\begin{equation}
\frac {r}{\zeta }=\frac {1-\frac {4}{(Bt_0)^2}+\frac {24}{(Bt_0)^4}}
{1-\frac {8}{(Bt_0)^2}}\approx 1+\frac {4}{(Bt_0)^2}+\frac {88}{(Bt_0)^4}>1.
\label{rz}
\end{equation}
The value of $r$ is always larger than $\zeta $. It exceeds $\zeta $ by about 
5$\%$ at 7 TeV. It is important that their ratio (\ref{rz}) depends on a single 
quite large parameter $Bt_0$. The dip near $t_0$ shifts (see the Table) 
at higher energies inside the region which at low energies traditionally 
belonged to the diffraction cone at $\vert t\vert <0.4$ GeV$^2$. The excess 
of $r$ over $\zeta $ becomes larger at higher energies 
because $(Bt_0)^2$ decreases as seen from the Table and Figure~\ref{fig:Bt0}. 
It can be used as a guide in comparing further results at higher energies.

Measuring the differential cross section, we get no information about the signs 
of real and imaginary parts of the amplitude. The kfk-model predicts that the 
imaginary part changes its sign near the dip of the differential cross section.
In principle, one can imagine another possibility to fit the differential 
cross section ascribing positive $f_I(s,t)\propto +\sqrt {d\sigma /dt}$, 
i.e. considering the positive tail of $f_I$. Unfortunately, it lacks
any explanation of the dip in the differential cross section as a zero of the 
imaginary part. This assumption was used in papers 
\cite{amal, dnec, ijmp, igse, white, mart}. Since the kfk-model claims 
to describe the tail of the differential cross section quite well, the proposal
of positive $f_I$ would mean that one should just subtract 
$\zeta [\vert t_0\vert, \infty ]$ from $\zeta [0,\vert t_0\vert ]$. The result 
is shown at the line $\zeta (\sigma )$. One concludes that, for the case of
the positive tail of the imaginary part, $\zeta $ in the 
unitarity condition (\ref{unit}) can be well approximated by $r$ shown in the
Table. The asymptotic predictions drastically change if $\zeta $ increases at 
higher than LHC energies and exceeds 1. The darkness of central collisions 
diminishes, the maximum absorption ($G=1$) moves to more peripheral values of 
the impact parameters $b$, and the interaction region looks like a toroid 
(see the Figure in \cite{ijmp, igse, white}). 

One reaches similar conclusions if just the exponential fit of the diffraction 
cone is used where the tail of $f_I$ is also positive but lower than 
for the $+\sqrt {d\sigma /dt}$-case. The predicted increase of $\zeta $
is somewhat slower than in the previous case but the asymptotic depletion
of the inelastic profile at $b=0$ is confirmed.

At present energies the values of $\zeta $ and $r$ differ by about 5$\%$.
Thus, the quadratically small decline of $G(s,0)$ from 1 claimed in \cite{mart} 
can not be noticed within the accuracy of experimental data.
The choice between different possibilities could be done if the drastic
change of the exponential shape  of the diffraction cone at higher energies
will be observed at higher energies. Some guide to that is provided in the 
kfk-model by the rapid motion of the positions of zeros to smaller values
of $\vert t_0\vert $ inside the former cone region with energy 
increase as seen from the Table and from the simplified treatment \cite{sur}. 
It is interesting to note that a slight decline from the exponentially 
falling $t$-shape of the diffraction peak was shown by the extremely precise 
data of the TOTEM collaboration \cite{tot3} already at 8 TeV. Is that the very 
first signature?

It was proposed \cite{jenk} to explain this decline as an effect of 
$t$-channel unitarity with the pion loop inserted in the Pomeron exchange graph. 
The $s$-channel "Cutkosky cut" of such a graph (its imaginary part) reproduces 
exactly the inelastic one-pion exchange graph first considered in \cite{drch}. 
Thus the external one-pion shell ascribed to inelastic peripheral interactions 
almost 60 years ago becomes observable at small $t$ of the elastic amplitude 
at 8 TeV.

Moreover, the energy behavior of the slope $B(s)$ looks somewhat different
from the logarithmic one prescribed to it by the Regge-poles. 
The transition between 2.76 TeV and 8 TeV data would ask for steeper 
dependence (see Figure~\ref{fig:TOTEM_B(s)}) and, 
therefore, for a non-pole nature of the Pomeron singularity.

\section{How different $b$-regions contribute to the $t$-structure of the 
elastic amplitude}

Here we can only rely on the kfk-model where the analytical expression for the 
amplitude in the $b$-representation (\ref{fkb}) is written. We demonstrate the
evolution with energy increase of the imaginary part of the amplitude as a
function of the transferred momentum in Figure 5 for the presently available 
energy 7 TeV (upper part) and for "asymptotically"  high energy $10^4$ TeV
(lower part) by the dash-dotted lines. It is clearly seen that the diffraction 
cone becomes much steeper and the zero moves to smaller transferred momenta at
higher energies.

To reveal the substructure, the contributions from different intervals of 
the impact parameter are also shown. We integrate over 
the three different regions of the impact parameters. They are quite naturally
dictated by the $b$-space shape of the amplitude (\ref{fkb}). The region of 
small impact parameters is up to values of $b\leq 2.5$ GeV$^{-1}\approx 0.5$ fm.
Here the exponential term is at least twice higher than $\Psi _K$ (see Fig. 7b
in Ref. \cite{kfk1}). Its contribution to $f_I$ is shown by the solid line in 
Figure 5. The region of large impact parameters (dotted line)
is considered at distances above 1 fm. The intermediate region (dashed line) 
lies in between them (2.5 - 6 GeV$^{-1}$). 
 
The most intriguing pattern is formed inside the diffraction cone.
The intermediate region of impact parameters $b$ (2.5-6 GeV$^{-1}$) contributes 
mainly there (especially for $0.15<\vert t\vert <0.4$ GeV$^2$) at 7 TeV. 
However quite sizable contributions appear from small impact parameters
at $0.07<\vert t\vert <0.15$ GeV$^2$ and, especially, from large impact 
parameters at very small $\vert t\vert <0.07$ GeV$^2$.

At $10^4$ TeV the peripheral region of large impact parameters strongly 
dominates at very small $\vert t\vert $ while the role of central 
interactions is diminished.

The remarkable feature of the kfk-model, its zero of $f_I$, appears due to 
compensations from small and medium impact parameters at 7 TeV for
$\vert t\vert \approx 0.5$ GeV$^2$. The peripheral region does not play
an important role in its existence. In contrast to it, namely peripheral
contribution is crucial at $10^4$ TeV. It is compensated by the sum from
the intermediate and central regions of the impact parameters at
$\vert t\vert \approx 0.1$ GeV$^2$. 

At larger $\vert t\vert $ the damped oscillatory pattern of contributions from 
intermediate and peripheral regions of $b$ dominates at all energies while
the steadily decreasing (with $\vert  t\vert $) share of central interactions
is rather unimportant. 

Let us remark that some special features of the elastic amplitude as a function 
of the impact parameters were discussed in Ref. \cite{shur}.
In the framework of the holographic approach, it was speculated that they
can be ascribed at intermediate values of $b$ to the exchange by the black 
hole tubes (BH) and at larger $b$ to the string holes (SH) (see Fig. 4 in Ref.
\cite{shur}). It is tempting to relate it to the substructure seen in Figure 5.
Namely these two regions contribute mainly to the diffraction cone at 7 TeV 
while lower impact parameters provide slowly changing background. Thus SH-tubes 
are in charge of the peripheral processes and therefore have larger sizes than 
BH. The steep behavior of the peripheral contribution at very low transferred 
momenta can lead to the non-exponential curvature of the diffraction cone 
noticed in experiment already at 8 TeV \cite{tot3, cs} and interpreted as a 
direct impact of the long-range one-pion exchange forces (see Ref. \cite{jenk}).
The role of peripheral processes (i.e. the long SH-tubes) increases with energy 
according to Figure 5. 

\begin{figure}[h]
\caption{The shapes of the imaginary part as a function of the transferred
momentum at 7 TeV (up) and $10^4$ TeV (below) are shown by the dash-dotted 
lines. The contributions to them from different impact parameters are also 
shown.}
\includegraphics[width=\textwidth]{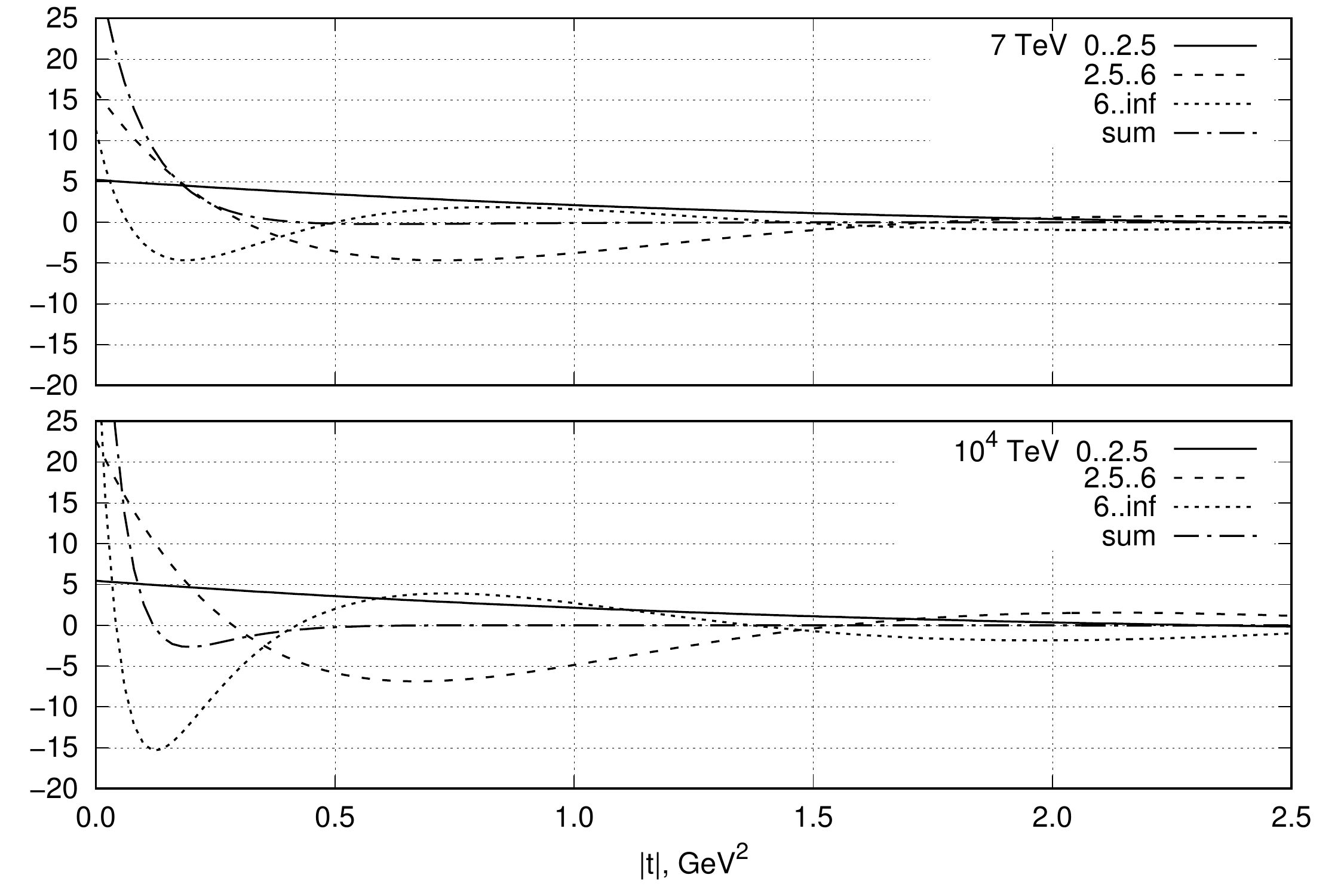} 
\label{fig:part_int}
\end{figure}

\section{Conclusions}

Elastic scattering of protons continues to surprise us through new experimental
findings. The implications of the peculiar shape of the differential cross 
section and the increase of the ratio of the elastic to total cross sections
with increasing energy are discussed above. These facts determine the 
spatial interaction region of protons. Its shape can be found from the unitarity 
condition within definite assumptions about the behavior of the elastic
scattering amplitude. The inelastic interaction region becomes more Black,
Edgier and Larger (BEL) in the energy range from ISR to LHC. Its further
fate at higher energies is especially interesting. 

There are two possibilities discussed in the paper. The previous tendency 
is conserved within the kfk-model. However the shape of the inelastic 
interaction region can evolve to the so-called TEH (Toroidal Elastic 
Hollow) regime with more gray region of central collisions. The sign of the 
imaginary part of the amplitude at rather large transferred momenta is 
responsible for the difference between these two possibilities. It is
negative in the kfk-model and positive for a different prescription.
That can not be found from present-day experiments and asks for 
phenomenological models. Probably, experiments with polarized protons
scattered at rather large angles of the Orear region can help.
The energy behavior of the ratio of elastic to total cross sections plays a 
crucial role. If observed, its rapid increase at higher energies would give 
some arguments in favor of the second possibility. 

More surprises could be in store for us from the evolution of the shape of 
the diffraction cone with the dip position moving to smaller transverse momenta 
and a drastic change of its slope as shown above. First signatures of the 
shape evolution can be guessed even at present energies from TOTEM-findings 
at 2.76 TeV and 8 TeV.

\medskip

{\bf Acknowledgments}

\medskip

I.D. is grateful for support by the RAS-CERN 
program and the Competitiveness Program of NRNU "MEPhI" ( M.H.U.).

\end{document}